\documentclass[11pt]{article} 

\usepackage{amsfonts}
\usepackage{amsbsy} 
\begin{document}
\newcommand{\hk}{HyperK\"{a}hler}
\newcommand{\be}{\begin{equation}}
\newcommand{\ee}{\end{equation}}
\newcommand{\mm}[1]{\ensuremath{\mu^{-1}(#1)}}
\newcommand{\ti}[1]{\ensuremath{\tilde{#1}}}
\newcommand{\bs}{\ensuremath{\boldsymbol{\omega}}}
\newcommand{\re}{\ensuremath{\mathbb{R}}}
\newcommand{\q}{\ensuremath{\mathbb{H}}}
%
\title{\hk \ Quotient Construction of\\ BPS Monopole Moduli Spaces.}
\author{G.W. Gibbons\thanks{e-mail:gwg1@damtp.cam.ac.uk} \ and\   
        P. Rychenkova\thanks{e-mail:pr201@damtp.cam.ac.uk} \\
        DAMTP, University of Cambridge, Silver Street \\ 
        Cambridge, CB3 9EW, U.K.}
\date{August 13, 1996}
\maketitle
\begin{abstract}
We use the \hk\ quotient of flat space to obtain some monopole moduli
space metrics in explicit form. Using this new
description, we discuss their topology, completeness and
isometries. We construct the moduli space metrics in the
limit when some monopoles become massless, which corresponds to
non-maximal symmetry breaking of the gauge group. We also introduce a
new family of \hk\ metrics which, depending on the ``mass parameter''
being positive or negative, give rise to either the asymptotic metric
on the moduli space of many $SU(2)$ monopoles, or to previously
unknown metrics. These new metrics are complete if one carries out the
quotient of a non-zero level set of the moment map, but develop
singularities when the zero-set is considered. These latter
metrics are of relevance to the moduli spaces of vacua of
three dimensional gauge theories for higher rank gauge
groups. Finally, we make a few comments concerning the existence of
closed or bound orbits on some of these manifolds and the
integrability of the geodesic flow.
\end{abstract}
\section{Introduction}
The purpose of this paper is to provide a simple and explicit construction, 
using the \hk \ quotient \cite{hkquo}, of some \hk \ metrics
which have been used to check the S-duality hypothesis in $N=4$ supersymmetric
Yang-Mills theory in four space-time dimensions \cite{lwy1,gaunt,lwy2,gary}
and which have also been recently applied to Yang-Mills theory in
three space-time dimensions \cite{intril}.

The use of the \hk\ quotient construction in this context is not in itself new,
but our treatment has the advantage that rather little machinery is necessary 
to obtain simple and tractable expressions for the metrics. Moreover, it 
allows us to analyse certain global properties of the manifolds, such as 
topology
and completeness, with comparatively little effort. We are able to make some
statements about the isometries and geodesics of these metrics which are not
obvious from the explicit forms given in \cite{lwy1,su2,lwy3}. Another
advantage of
this approach is that it greatly simplifies the analysis of the
metrics on monopole moduli spaces when some of the monopoles become massless. 
The construction permits an easy examination of singularities and how
they may be resolved by changing the level sets of the moment map.

The plan of the paper is as follows. In section~\ref{general} we review the
\hk\ quotient construction and, in particular,  how starting from a flat \hk\ 
structure on $\q^{m+d}\equiv \re^{4(m+d)}$, we can obtain a $4m-$dimensional
\hk\ metric, using a $d-$dimensional subgroup $G$ of the Euclidean group
$E(4m+4d)$. In all our examples there is a tri-holomorphic action of the torus
group $T^m = U(1)^m$ on $\q^{m+d}$ which commutes with the action of $G$ and 
therefore induces a tri-holomorphic $T^m$ action on the quotient. The general 
local forms of the metrics admitting a tri-holomorphic torus action have been
written down previously by Lindstr\"{o}m \& Ro\v{c}ek \cite{linroc} and
Pedersen \& Poon 
\cite{poon}, up to the solution of a set of linear partial differential 
equations. Further work on this type of metrics is contained in
\cite{goto}. While in the most of the previous work on monopole moduli
spaces the solutions
were deduced from an asymptotic (Li\'{e}nard-Wiechert) analysis of interactions
of the monopoles, the quotient construction gives the required solutions 
directly. An advantage of this construction is that one can also
imagine taking the limit of large $m$, which  
in some cases would correspond to many monopoles and in others to a gauge group
$SU(m+2)$ for large $m$; this can be done naturally in the \hk\ quotient 
setting.

Section~\ref{example} is devoted to giving some explicit examples.
We obtain the Lee-Weinberg-Yi metric on $\re^{4m}$ \cite{lwy1} which
is the relative 
moduli space of $m+1$ distinct fundamental monopoles with gauge group $SU(m+2)$
broken down to $U(1)^{m+1}$. As we shall show, the Lee-Weinberg-Yi metric is
determined by $m$ linearly independent vectors in $\re^m$ whose matrix of
inner products gives the reduced mass matrix of the monopoles.
As well as the Lee-Weinberg-Yi metric, we use this technique to construct the
Calabi metrics on $T^*(\mathbb{CP}^m)$, the Taubian-Calabi metrics on $\re^
{4m}$ and the cyclic ALE and ALF four-metrics. They are relevant for the 
limiting cases of zero and infinite monopole mass which are the subject of 
section~\ref{massmon}. As the last and probably most elaborate example of 
section~\ref{example} we construct an apparently new class of metrics which 
include as a special case a positive mass parameter version of the asymptotic
metric on the moduli space of $m+1$ identical $SU(2)$ monopoles \cite{su2}.
The general metric is complete and depends on $\frac{1}{2}m(m-1)\ $ 
three-vectors $\boldsymbol{\zeta}_{ab}$. When all
$\boldsymbol{\zeta}_{ab}=0$ the manifold has singularities for both
positive and negative mass parameters. The topology
of these metrics is rather complicated and we defer its consideration until a 
later publication. Metrics of this type have recently figured in
studies of gauge theory in three dimensions \cite{intril}.

In section~\ref{massmon} we discuss the limiting forms of the Lee-Weinberg-Yi
metric when the masses of one or more monopoles either vanish or become
infinite. In our formalism this is equivalent to one or more of the $m$ 
vectors describing the metric becoming infinitely large or vanishing 
respectively, but one can still perform the quotient in these cases. If all but
two monopoles become massless, and $SU(m+2)$ breaks to $SU(m)\times
U(1)^2$, the
metric on the relative moduli space is the Taubian-Calabi metric. It was 
obtained in \cite{lwy3} but not recognised as such. As we show, if $SU(m+2)$ 
is broken to $SU(m+2-k)\times U(1)^k$ the construction works just as well. At
the end of this section we turn to the opposite limit when one or more 
monopoles become infinitely massive. If only one mass remains finite we obtain
the $m-1$ centre ALF metric.

Finally in section~\ref{geodesic} we discuss some aspects of the geodesic 
motion. In particular we use the quotient construction to show that neither the
Lee-Weinberg-Yi nor the Taubian-Calabi metrics admit closed or even bound 
geodesics. We also make some remarks concerning the integrability of the
geodesic flow. The last section contains a few concluding comments.

\section{\hk \ Quotients}  \label{general}

In this section we shall recall some properties of the \hk \ quotient
construction \cite{hkquo} that we shall need and establish our
notation. If $\{\mathcal{M},g,I,J,K\}$ is a \hk \ manifold and $G$ a Lie group
with Lie algebra $\mathfrak{g}$ which acts on $\mathcal{M}$ preserving the 
\hk \ structure, there will be an associated moment map $\mu :\mathcal{M} 
\rightarrow \re^3 \otimes \mathfrak{g}^*$. 

If we pick an element $\zeta \in \mathcal{Z}$, the centre of $\mathfrak{g}^*$,
i.e. the invariant element of $\mathfrak{g}^*$ under the co-adjoint
action, then
\[ X_\zeta := \mm{\zeta}/G \]
is also a \hk \  manifold.                   
If $G$ is compact and acts freely on $\mm{\zeta}$ and
$\mathcal{M}$ 
is complete then $ X_\zeta$ is also complete\footnote{In
the examples of this paper $\mathcal{M}$ and $G$ are both non-compact,
so to check the completeness of the quotient manifold one has to check
both the freedom of $G$-action and the behaviour at infinity.}. If $G$ does 
not act freely on $\mm{\zeta}$ then  $ X_\zeta$ will, in general, have 
singularities.  

The manifold $\mathcal{M}$ may also admit another group $K$ whose
action preserves
the \hk \ structure on $\mathcal{M}$ and commutes with the action of
$G$. Then $K$ will descend to  $ X_\zeta$ as a group of 
tri-holomorphic isometries. 

Note that, at least locally, everything  we have said about Riemannian
metrics carries over in a straightforward way to metrics of signature 
$(4p,4q)$. This comment will be of use later but, unless stated
otherwise, we shall be concerned with the usual positive definite
case.

In this paper we consider the case
\[ \mathcal{M} = \re^{4n} \cong \q^n \] 
with its standard flat metric. The group $G$ is a (typically non-compact)
subgroup of the Euclidean group $E(4n)$ which preserves the standard
\hk \ structure.

In order to establish notation we consider the case $n=1$. We may
identify points $(w,x,y,z) \in \re^4$ with a quaternion $q\in \q$:
\be
q = w+ ix+jy+kz.
\ee    
The metric is
\be\label{Hflat}
ds^2 = dq \, d\bar{q},
\ee
where $\bar{q} = w-ix-jy-kz$ and the three K\"{a}hler forms are
\be
- \frac{1}{2} dq \wedge d\bar{q} = i \omega_I + j \omega_J + k\omega_K.
\ee

The \hk \  structure is invariant under real translations
\be
q \rightarrow q + t ,  t \in \re,
\ee
 and right multiplications by unit quaternions $\cong SU(2)$
\be
q \rightarrow q \, p,
\ee
$p\bar{p} = 1$. By contrast left multiplication by a unit quaternion 
\[ q \rightarrow p\, q\]
is an isometry of the metric (\ref{Hflat}) but rotates the three
K\"{a}hler forms. This may also be seen by noting that since the
metric is flat we may identify the tangent space with $\q$, then the
complex structures $I,J,K$ act on $\q$ by left multiplication by
$i,j,k$. The action of $I,J,K$ thus commutes with right
multiplication.

The moment map for real translations is
\be
\mu = \frac{1}{2}(q-\bar{q}).
\ee
The one parameter family of right multiplications 
\be \label{u1action}
q \rightarrow q \, e^{it},\ \ t \in (0,2\pi] 
\ee
has moment map\footnote{The awkward factor of $2$ appears in
the moment map so as to enable us to make further definitions more natural.}
\be \label{rotationmm}
\mu =\frac{1}{2} q i \bar{q}.
\ee
Away from the origin, $q = 0$, the $U(1)$ action (\ref{u1action}) is
free. The moment map (\ref{rotationmm}) allows us to identify the orbit
space with $\re^3$, the origin corresponding to the fixed point set
$q=0$. The moment map (\ref{rotationmm}) thus defines a Riemannian
submersion $\re^4 \setminus 0 \rightarrow \re^3 \setminus 0$ whose
fibres are circles $S^1$.

For later purposes it will prove useful to
express the flat metric (\ref{Hflat}) in coordinates adapted to the
submersion. Any quaternion may be written as 
\be
q= a e^{i \psi/2},
\ee
where the real coordinate $\psi \in (0, 4\pi ]$ and $a$ is pure
imaginary, $a = -\bar{a}$. Then the $U(1)$ action is given by 
\be
\psi \rightarrow \psi + 2t.
\ee
The moment map (\ref{rotationmm}) defines three cartesian coordinates 
$\mathbf{r}$ by
\be
\mathbf{r} = qi\bar{q} = ai \bar{a} = -aia.
\ee
A short calculation reveals that the flat metric (\ref{Hflat}) in
coordinates $(\psi, \mathbf{r})$ becomes
\be \label{flatmet}
ds^2 = \frac{1}{4} \Bigl(\frac{1}{r}d\mathbf{r}^2 + r(d\psi +\bs. 
d\mathbf{r})^2 \Bigr),
\ee
with $r =|\mathbf{r}| $ and
\[
\mathrm{curl} \, \bs = \mathrm{grad} \, (\frac{1}{r}),
\]
and where the $\mathrm{curl}$ and $\mathrm{grad}$ operations are taken with 
respect to flat euclidean metric on $\re^3$ with cartesian coordinates
$\mathbf{r}$.
The metric (\ref{flatmet}) is singular at $r=0\equiv q=0$ but this is
merely a coordinate artefact arising from the $U(1)$ action
(\ref{u1action}) having a fixed point there. Away from the fixed point
the metric (\ref{flatmet}) is defined on the standard Dirac circle
bundle over $\re^3\setminus 0$ and the horizontal one-form $(d\tau +
\bs . d\mathbf{r} )$ defines the standard Dirac
monopole connection.

In the next section we shall use the form of the metric (\ref{flatmet})
for $\re^4$ and the Hyperk\"{a}hlerian $U(1)$ action (\ref{u1action}) to
construct various \hk \ quotients of $\re^{4n}$.

\section{Explicit quotient constructions}   \label{example}

In this section we will give examples of some \hk \ metrics constructed from
flat space. Before proceeding we note that all of our examples of
$4m-$dimensional \hk \ metrics admit a tri-holomorphic $T^m$ action
which implies 
that locally the metrics may be cast in the form \cite{poon}:
\be   \label{genmet}
ds^2 = \frac{1}{4}G_{ab} d{\bf r}_a . d{\bf r}_b +\frac{1}{4} G^{ab} 
(d \tau _a +\bs_{ac} . d {\bf r}_c)(d \tau _b +\bs_{bd} . d {\bf r}_d),
\end{equation} 
where $a,b=1 \ldots m$,  $G^{ab}$ is the inverse of $G_{ab}$, and the Killing 
vector fields $\partial/\partial \tau _a$ generate the $T^m$ action. Unless 
otherwise stated, we will assume Einstein summation convention. 

The matrix $G_{ab}$ and the one-form components $\bs_{ab}$
satisfy certain {\it linear}  equations which have the property that given a 
solution for the matrix $G_{ab}$ one may determine the one forms up to gauge
equivalence. Thus to identify a metric of this form we need only to calculate
$G_{ab}$. We shall use this fact later to relate metrics obtained by the quotient construction to previously known forms. From the discussion 
in the previous section  the reader should be able to recognise 
that if $r_a = |\mathbf{r}_a|$ and 
\be
G_{ab} = { \delta_{ab} \over r_a}
\ee
then one obtains the flat metric on ${\mathbb H}^m$.

Because the torus action is tri-holomorphic it has an associated 
moment map which in the present case is given (up to a scalar multiple) by 
\be
\mu = {\bf r}_a,
\ee
which may be used to parameterise the space of orbits of the $T^m$ action.
Away from the degenerate orbits $ \Delta$ of the torus 
action the map from $X_\zeta$ to $\re^{3m}$ is therefore a Riemannian
submersion. This defines a torus bundle over $\re^{3m} \setminus \Delta$
with a connection whose horizontal one-forms are just
\be
A^a = G^{ab} (d \tau _a + \bs_{ac} . d {\bf r} _c) .
\ee

Because  \hk \  metrics are necessarily  Ricci flat, it follows 
from the Killing's equations that the exact 2-forms
\begin{equation}
F^a = dA^a
\end{equation}
are co-closed
\begin{equation}
d^{*}F^ a=0,
\end{equation}
and hence harmonic. By taking wedge products we obtain a useful supply
of even-dimensional exact and co-closed harmonic forms. This
construction yields the middle-dimensional Sen form for fundamental
monopoles \cite{gary}.

\subsection{Taub-NUT space}  \label{taubn}
The prototype case of the constructions we are interested in is that
of Taub-NUT space. We will describe this in great detail and omit explicit
calculations for the later cases. Choose
\be
\mathcal{M} = \q \times\q
\ee
with quaternionic coordinates $(q,w)$. Let $G$ be $\re$, $t \in \re$, with action
\be \label{tnaction}
(q,w) \rightarrow (qe^{it}, w + \lambda t), \ \lambda \in \re,
\ee
and moment map
\begin{eqnarray} 
\mu & = &{1 \over 2} qi\bar{q} + \frac{\lambda}{2}(w-\bar{w}) \\
    & = & {1 \over 2}\mathbf{r} + \lambda \mathbf{y},  \nonumber
\end{eqnarray}
where $w = (y + \mathbf{y}) , y \in \re$. The flat metric on
$\mathcal{M}$ is
\be
ds^2= \frac{1}{4}\biggl(\frac{1}{r}d\mathbf{r}^2 + r(d\psi + \bs . 
d\mathbf{r})^2 \biggr) + dy^2 + d\mathbf{y}^2.
\ee
The action (\ref{tnaction}) corresponds to $(\psi,y)\rightarrow
(\psi+2t,y+\lambda t)$, which leaves $\tau = \psi - 2y/\lambda$
invariant. We set, without loss of generality, 
\be
\zeta = 0.
\ee
On the five-dimensional intersection of the three level sets
$\mm{0}$ one has $\mathbf{y} = - \mathbf{r}/2\lambda$ so the
induced metric is:
\be \label{tn1}
ds^2= \frac{1}{4}\biggl(\frac{1}{r}d\mathbf{r}^2 +r(d\tau +\frac{2}{\lambda}dy
+ \bs . 
d\mathbf{r})^2 \biggr) + dy^2 +\frac{1}{4 \lambda^2} d\mathbf{r}^2. 
\ee
The metric on $\mm{0}/\re$ is obtained by projecting orthogonally to the
Killing vector field $\partial / \partial y$. Completing the square in
(\ref{tn1}) gives
\begin{eqnarray*}
ds^2 & = & \frac{1}{4} \biggl(\frac{1}{r} + \frac{1}{\lambda^2} \biggr) 
           d\mathbf{r}^2
           +\frac{1}{4}\biggl( \frac{1}{r} + \frac{1}{\lambda^2} \biggr)^{-1}
           \biggl( d\tau + \bs. d\mathbf{r} \biggr)^2 \\ 
     &   & \mbox{} + \biggl( \frac{r}{\lambda^2}+ 1 \biggr) \biggl(dy + 
           \frac{r\lambda}{2}\frac{(d\tau +\bs 
           . d\mathbf{r})}{(\frac{r}{\lambda^2}+ 1)}\biggr)^2
\end{eqnarray*}
Thus the metric on $\mm{0}/\re$ is 
\be \label{tnmet}
ds^2 = \frac{1}{4} \biggl(\frac{1}{r} + \frac{1}{\lambda^2}\biggr) 
        d\mathbf{r}^2 +\frac{1}{4}\biggl(\frac{1}{r} + \frac{1}{\lambda^2}
         \biggr)^{-1} (d\tau + \bs . d\mathbf{r})^2.
\ee
This is the standard form of the Taub-NUT metric with positive ``mass
parameter''. It becomes singular at $r=0$ but since, as is easily seen,
$\re$ acts freely on $\mm{0}$ this is a coordinate singularity. To
obtain global coordinates on the quotient space note that on $\mm{0}$,  
$\mathbf{y} = -qi\bar{q}/2$ and the $\re$ action shifts $y$, so we may
set $y = 0$. Thus $q$ serves as a global coordinate and we can see that
topologically the Taub-NUT metric is equivalent to $\re^4$. Also if 
$\lambda \rightarrow \infty$ Taub-NUT metric (\ref{tnmet}) degenerates 
to a flat metric on $\re^4$ of the form (\ref{flatmet}).

The Taub-NUT metric with negative mass parameter may be obtained in an
analogous way but now starting with the flat metric of signature
$(4,4)$:
\[ ds^2 = dq \, d\bar{q} - dw \,d\bar{w}. \] 
Following the steps above yields the same metric (\ref{tnmet}) but with 
$\lambda^2$ replaced by $-\lambda ^2$.

The $\re$ action (\ref{tnaction}) commutes with the $U(1)$ action:
\[ (q,w) \rightarrow (q e^{i\alpha},w) \]
which descends to the Taub-NUT metric as the tri-holomorphic action
$\tau \rightarrow \tau + 2\alpha$. In addition the action of the unit
quaternions \[ (q,w) \rightarrow (p\,q, p\, w \, \bar{p}), \]
$p\bar{p} = 1$, the $SU(2)$ action, commutes with both of the previous actions
and leaves $\mm{0}$ invariant. Therefore the full isometry group of the Taub-
NUT metric is $U(2)$.
\subsection{The Lee-Weinberg-Yi metric} \label{LWY}
The simplest generalisation of Taub-NUT metric is perhaps the metric on 
the relative moduli
space of distinct fundamental monopoles when the gauge group is
$SU(m+2)$ broken down to its maximal torus $U(1)^{m+1}$ \cite{lwy1,mur}. The 
case $m=1$ coincides with Taub-NUT metric which is the exact 
metric on the relative moduli space of $SU(3)$ fundamental monopoles 
\cite{conn,gaunt,lwy2}.
We take
\be
\mathcal{M} = \q^m \times \q^m
\ee 
with coordinates $(q_a, w_a)$, $a = 1,\ldots , m$, and $G = \re^m
= (t_1,\ldots,t_m)$ with action
\begin{eqnarray} \label{lwyaction}
q_a & \rightarrow & q_a \, e^{it_a} \ \ \   (\mathrm{no\  sum\  over}\ a), \\
w_a & \rightarrow & w_a + \lambda_a^{\ b} t_b.   \nonumber
\end{eqnarray}
The action of G commutes with the tri-holomorphic action of $K = T^m =
U(1)^m = (\alpha_1, \ldots, \alpha_m)$ given by
\begin{eqnarray} \label{lwytaction}
q_a & \rightarrow & q_a \, e^{i\alpha_a} \ \ (\mathrm{no\ sum\  over}\ a), \\
w_a & \rightarrow & w_a. \nonumber
\end{eqnarray}
The moment maps of the $\re^m$ action are
\be  \label{lwymm}
\mu_a = {1 \over 2} q_a \, i\, \bar{q}_a  + \frac{1}{2}\lambda_a^{\ b} (w_b
-\bar{w}_b),
\ee
where the $m \times m$ real matrix $\lambda_a^{\ b}$ is taken to be
non-singular. The Lee-Weinberg-Yi metric is then the induced metric on
$\mm{0}/\re^m$. The zero set of the moment maps is given by
\be  \label{lwyzero}
\mu_a = {1\over 2}\mathbf{r}_a + \lambda_a^{\ b} \mathbf{y}_b = 0,
\ee
where $\mathbf{r}_a =q_a\, i \, \bar{q}_a $ and $\mathbf{y}_a =
1/2(w_a - \bar{w}_a)$.  
A short calculation shows that the metric on the quotient is
\begin{eqnarray} \label{lwymet}
ds^2     & = & \frac{1}{4}G_{ab}d\mathbf{r}_a. d\mathbf{r}_b+\frac{1}{4}G^
               {ab}(d\tau_a +\bs(\mathbf{r}_a). d\mathbf{r}_a)(d\tau_b+\bs
               (\mathbf{r}_b) . d\mathbf{r}_b), \\
G_{ab}   & = & \frac{\delta_{ab}}{r_a} +\mu_{ab},  \nonumber
\end{eqnarray}
where
\be  \label{massmat}
\mu_{ab} = (\nu^t \, \nu)_{ab}
\ee
\[ \nu \equiv \lambda^{-1}, \]
and
\be
\mathrm{curl}_c \, \bs(\mathbf{r}_a) = \mathrm{grad}_c\, (\frac{1}{r_a}).
\ee
The tri-holomorphic action is generated by $\partial/\partial\tau_a$. 
Condition (\ref{lwyzero}) is invariant under the action of the unit quaternions
\begin{eqnarray}
q_a & \rightarrow & p\, q_a,  \\
w_a & \rightarrow & p \,w_a \bar{p}. \nonumber
\end{eqnarray}
The metric on $\mm{0}/\re^m$ is thus invariant under $SU(2)$ and a
tri-holomorphic action of $T^m$ corresponding to (\ref{lwytaction}).
 As remarked above: to specify a $4 m\,$-dimensional
\hk \ metric with a tri-holomorphic $T^m$ action it suffices to
specify the $m \times m$ matrix $G_{ab}$ as a function of the
$\mathbf{r}_a$'s. The remaining parts of the metric may then be
deduced directly. In the present case $G_{ab}$ depends on the matrix
$\mu_{ab}$. Physically $\mu_{ab}$ is, up to a constant factor, the
reduced mass matrix in the centre of mass frame of the
monopoles. Geometrically it is related to the matrix of inner products
of the $m$ linearly independent translation vectors defining the
$\re^m$ action. The translation vectors $v^{(b)}$,$v \in \re^m$,
$b=1,\ldots,m$ have components:
\[ (v^{(b)})_a = \lambda_a^{\ b}. \]
Thus \[ g(v^{(a)}, v^{(b)}) = (\lambda^t \,\lambda)^{ab} = \mu^{ab}. \]
If one thinks of the vectors $v^{(b)}$ as defining a lattice
$\Lambda$ in $\re^m$ with metric $ g(v^{(a)},v^{(b)})$ then $\mu_{ab}$
 are the components of the metric on the reciprocal lattice
$\Lambda^*$. 

As long as the matrix $\lambda_a^{\ b}$ is invertible we may use
(\ref{lwyzero}) to eliminate $\mathbf{y}_b$ in favour of
$\mathbf{r}_a$ on $\mm{0}$. It then follows, as it did for the Taub-NUT
case, that the $m$ quaternions $q_a$ will serve as global coordinates
for the Lee-Weinberg-Yi manifold which is therefore homeomorphic to
$\re^{4m}$. The completeness of the metric follows from the fact
that $\re^m$ acts freely on $\mm{0}$ and an examination of the metric
near infinity.

If the matrix $\lambda_a^{\ b}$ becomes singular or diverges, i.e. if
the translation vectors cease to be linearly independent  or become
infinite, we are led to various degenerate cases associated with the
reduced mass matrix $\mu_{ab}$ dropping in rank. Physically this is
associated with enhanced symmetries due to appearance of massless
monopoles. We shall discuss this in more detail in section~\ref{massmon}.

Above we have considered monopoles of a specific gauge group
$SU(m+2)$, in fact this
construction may be easily generalised for any semi-simple group of rank 
$m+1$. In notation of \cite{lwy1}, $\lambda_a$'s (not to be confused with the
translation matrix $\lambda_{ab}$) are essentially inner 
products between the simple roots of the Dynkin diagram for the gauge group.
One replaces the flat metric on $\mathcal{M}$ by 
\[ \sum (\lambda_a dq_a d\bar{q}_a + dw_a d\bar{w}_a), \]
the K\"{a}hler forms by
\[-\frac{1}{2}\lambda_a
dq_a\wedge d\bar{q}_a-\frac{1}{2}dw_a\wedge d\bar{w}_a,\]
and the action (\ref{lwyaction}) by
\begin{eqnarray}
q_a & \rightarrow & q_a e^{it_a\lambda_a}\ \ \ (\mathrm{no\ sum}),\nonumber \\ 
w_a & \rightarrow & w_a + \lambda_a^{\ b}t_b.\nonumber
\end{eqnarray}
The form of the moment map (\ref{lwymm}) is unchanged.  
\subsection{The Calabi metrics on $T^{*}(\mathbb{CP}^m)$}
The construction of the Calabi metric on $T^{*}(\mathbb{CP}^m)$ is perhaps the
oldest of the \hk \ constructions \cite{calabi}. We describe it here
because we shall need it later. We choose
\be
\mathcal{M}= \q^{m+1}
\ee
with coordinates $q_a\,$,$a=1, \ldots,m+1$, and $G = U(1)$ with action
\be \label{caction}
q_a \rightarrow q_a \, e^{it}, \ \ \  t\in (0,2\pi]
\ee
and moment map
\be
\mu = \frac{1}{2}\sum q_a \, i\, \bar{q}_a =\frac{1}{2}\sum \mathbf{r}_a,
\ee
where $\mathbf{r}_a =q_a \, i\, \bar{q}_a$. The level sets of the
moment map $\mm{\zeta}$ are given by 
\be
\mu = \frac{1}{2}\sum \mathbf{r}_a  = \boldsymbol{\zeta}
\ee
where the $3-$vector $\boldsymbol{\zeta}$ must be non-vanishing if the 
action (\ref{caction}) is to be free. Let us make the following redefinition 
to make the formulas tidier:
\[ \boldsymbol{\zeta} = \frac{1}{2}\mathbf{x}.\]
Then the potential function $G_{ij}$ in (\ref{genmet}) is:
\begin{eqnarray}
G_{ii} & = & \frac{1}{|\mathbf{x} - \sum \mathbf{r}_i|}+\frac{1}{r_{i}} \\
G_{ij} & = & \frac{1}{|\mathbf{x} - \sum\mathbf{r}_i|},\ \ i \neq j \nonumber
\end{eqnarray}
and $i,j = 1, \ldots, m$.
The action (\ref{caction}) commutes with tri-holomorphic action of $SU(m+1)$  
given by 
\be
q_a \rightarrow q_a\, U_{ac},
\ee
where $U_{ac}$ is a $(m+1)\times(m+1)$ quaternion valued matrix with no $j$ or
$k$ components satisfying
\[
U_{ac}\bar{U}_{ab} = \delta_{cb},
\]
\[
\mathrm{det}\, U = 1. \]
Left multiplication by a unit quaternion 
\[ q_a \rightarrow p \, q_a \]
induces rotation of $\mathbf{r}_a$'s. If we choose $p$ such that this
is an $SO(2)$ rotation about the $\boldsymbol{\zeta}$ direction it will leave
$\mm{\zeta}$ invariant. Such an $SO(2)$ action will preserve a single
complex structure.

Thus the Calabi metric is invariant under the effective action of
$U(m+1)/
\mathbb{Z}_{m+1}$ of which $SU(m+1)/\mathbb{Z}_{m+1}$ acts tri-holomorphically.
With respect to a privileged complex structure we have a holomorphic effective
action of $U(m+1)/\mathbb{Z}_{m+1}$. The principal
orbits are of the form $U(m+1)/U(m-1) \times U(1)$. There is a degenerate orbit
of the form $U(m+1)/U(m)\times U(1) \cong \mathbb{CP}^m$ corresponding to the 
zero section of $T^{*}(\mathbb{CP}^m)$. 

A recent theorem of Swann and Dancer \cite{swan} shows that the Calabi
metric is the unique complete \hk \ metric of dimension greater than
four which is of cohomogeneity one\footnote{i.e. a manifold on which
the generic or principle orbit of the isometry group has real
codimension one}.
If $\boldsymbol{\zeta}=0$ the metric becomes incomplete -- it has an orbifold 
singularity at $q = 0$.
\subsection{The Taubian-Calabi metrics}  \label{taubcal}
The name Taubian-Calabi is due to \cite{tcalabi}.
Take
\be
\mathcal{M} = \q^{m}\times \q
\ee
with coordinates $(q_a,w)$,$\ a=1,\ldots,m$, and $G = \re$ with action
\begin{eqnarray}
q_a & \rightarrow & q_a \, e^{it},  \\
w   & \rightarrow & w+t, \nonumber
\end{eqnarray}
$t\in \re$. The moment map is
\be
\mu = {1 \over 2}\sum q_a \, i \bar{q}_a + \frac{(w-\bar{w})}{2}.
\ee
Without loss of generality $\mm{0}$ is given by 
\be
{1\over 2}\sum \mathbf{r}_a + \mathbf{y} = 0,
\ee
where as before $\mathbf{r}_a = q_a \, i \bar{q}_a$ and $\mathbf{y} =
1/2(w-\bar{w})$. There is a $T^m$ action on the $\q^m$ factor commuting with
$G$, therefore the metric is of the general form (\ref{genmet}) with metric 
components $G_{ab}$:
\begin{eqnarray}  \label{tcal}
G_{aa} & = & 1 + \frac{1}{r_{a}}  \\
G_{ab} & = & 1, \ \ \ a \neq b.  \nonumber
\end{eqnarray}
$X_{0}$ has a tri-holomorphic right action of $U(m)/\mathbb{Z}_m$ and a 
left action of $SU(2)$. The total isometry group of the Taubian-Calabi
metric is
then $U(m) \times SU(2)$ up to a discrete factor. The principle orbits of 
$U(m)$ are $U(m)/U(m-2)$ which are $(4m-4)-$dimensional. The left action of 
$SU(2)$ rotates the $q_a$'s and therefore $\mathbf{r}_a$'s but leaves
invariant the phase of
the $q_a$'s, thus it increases the dimension of a principle orbit by two. We 
conclude that the principle orbits of the Taubian-Calabi metric are of 
codimension two. As in the case of Lee-Weinberg-Yi metric, the $q_a$'s serve 
as global coordinates, and we get a complete metric on $\re^{4m}$. Setting 
$m = 1$ gives Taub-NUT metric. We will give a detailed discussion of the $m=2$
case in the next section.

In addition to continuous symmetries the Taubian-Calabi metrics admit
many discrete symmetries. There are $m$ reflections
$R_a:\,q_a\rightarrow -q_a$ and $S_m$ permutation group on $m$
letters, both acting tri-holomorphically. Their fixed point sets are
totally geodesic and \hk. In this way one sees that the
$4m-$dimensional Taubian-Calabi manifold contains as a totally
geodesic submanifold the $4n-$dimensional Taubian-Calabi manifold, for
$n < m$.
\subsection { The cyclic ALE metrics}
These metrics constitute an example of gravitational multi-instantons, 
complete $4-$dimensional solutions to vacuum Einstein equations, that were 
originally written down and discussed in \cite{minst}.
We take
\be
{\cal M} = {\mathbb H} ^m \times \q
\ee
with coordinates $(q_a,q)$, $a = 1, \ldots ,m$, and $G=T^m = 
(t_1,\ldots,t_m)$ with action
\begin{eqnarray} 
q_a & \rightarrow & q_a e^{it_a } \ \ \  (\mathrm{no\  sum})  \\
q   & \rightarrow & q e^{i \sum t_a }.  \nonumber
\end{eqnarray}
The moment maps for this action are
\be
\mu_a ={1 \over 2} ( q_a\, i\,\bar{q}_a + q\,i\,\bar{q}).
\ee
If ${\bf r}_a= q_a i\,\bar{q}_a$ and ${\bf r}= q\,i\,\bar{q}$, then 
$\mm{\zeta}$ is given by
\be
\frac{1}{2}{\bf r}_a= \boldsymbol{\zeta} _a - \frac{1}{2}\mathbf{r}.
\ee
For future convenience define $\boldsymbol{\zeta}_a = 1/2 \mathbf{x}_a$, then
the level sets of the moment map are:
\[ {\bf r}_a= \mathbf{x}_a - \mathbf{r}, \]
and the metric on $X_\zeta$ takes the multi-centre form 
\be   \label{multi}
ds^2 = { 1 \over 4} V d {\bf r}^2 + { 1\over 4} V ^{-1} (d\tau +\bs 
. d{\bf r} )^2, 
\ee
with
\begin{equation}
V = \frac{1}{r} + \sum { 1 \over |{\bf r}-\mathbf{x}_a |},
\end{equation}
and
\be 
{\rm curl} \, \bs = {\rm grad}\, V.
\ee
Because
\begin{equation}
{\bf r}_a - {\bf r} _b= \mathbf{x}_a - \mathbf{x}_b,
\end {equation}
we require $ \boldsymbol{\zeta}_a \ne \boldsymbol{\zeta} _b $, $\forall a,b$ 
in order that the action of $T^m$ be free. Note that the case $m=1$ coincides 
with the Eguchi-Hanson metric on $T^*(\mathbb{CP} ^1) $ which is the first of 
the Calabi series of metrics. The isometry group of the multi-instanton 
metrics is just $U(1)$ unless all the centres lie on a straight line
in which case there is an extra $U(1)$ symmetry. 
\subsection { The cyclic ALF metrics}  \label{alf}
These are the so-called multi-Taub-NUT metrics constructed by Hawking in 
\cite{mtn}. Take 
\begin{equation} 
{\cal M} = {\mathbb H}^m \times {\mathbb H}
\end{equation}
with coordinates $(q_a, w)$, $a=1,\ldots,m$, and $G=\re^m$ with moment map:
\begin{equation} \label{alfmm}
\mu_a= {1\over 2} {\bf r}_a + {\bf y},
\end{equation}
where ${\bf r}_a = q_a\,i\,\bar{q}_a $ and ${\bf y} = {(w-\bar{w})/2}$. As 
before make the following redefinitions:
\[ \mathbf{y} = \frac{1}{2} \mathbf{r}, \ \ 
\mathbf{x}_a = \frac{1}{2} \boldsymbol{\zeta}_a. \]  
The metric on $X_\zeta$ is again of multi-centre form (\ref{multi}) but this 
time with
\be  \label{alfmet}
V = 1+\sum \frac{1}{|\mathbf{r} - \mathbf{x}_a|}.
\ee
We must require $ \boldsymbol{\zeta}_a \ne \boldsymbol{\zeta} _b$ 
to avoid orbifold singularities at the coincidence points, that is when two or
more centres coincide. The ordinary Taub-NUT metric is the $m=1$ case.  
\subsection {The asymptotic metric on the moduli space of $SU(2)$
monopoles and its variations}
A more complicated example which also generalises Taub-NUT space corresponds
 to a form of the asymptotic metric on the moduli space of $m$ $ SU(2)$ BPS 
monopoles \cite{su2}. We begin by constructing the analogues of the Taub-NUT
metric with positive mass parameter and then go on to consider the analogue of
the Taub-NUT metric with negative mass parameter. It is this latter case which 
applies to the behaviour of $SU(2)$ BPS monopoles at  large separation.

We choose
\begin{equation} 
{\cal M} = \q ^ { { 1\over 2} m(m-1)} \times \q^m
\ee
with coordinates $( q_{ab}, w_a)$ , $a=1,\dots ,m$, $a<b$. The group $G$ is 
taken to be $\re ^ { { 1\over 2} m(m-1) } = (t_{ab})$ with action
\begin{eqnarray}
q_{ab} & \rightarrow & q_{ab} e^{i t_{ab}}, \\
w_a    & \rightarrow & w_a + \sum_{c} t_{ac}, \nonumber
\end {eqnarray}
where $t_{ac} = - t_{ca}$ for $c < a$.
The moment maps are
\begin{equation}
\mu _{ab} = {1 \over 2}{\bf r} _{ab} - ({\bf y}_a  - {\bf y}_b ),
\end{equation}
where $\mathbf{r}_{ab} = q_{ab}\,i\,\bar{q}_{ab}$ and $\mathbf{y}_a = 
(w_a - \bar{w}_a)/2$, then $\mm{\zeta}$ is given by
\begin{equation} \label{gmzero}
{1\over 2}{\bf r} _{ab}= {\bf y}_a - {\bf y}_b + \boldsymbol{\zeta}_{ab}.
\end{equation}
Using (\ref{gmzero}) one may eliminate the ${\bf r}_{ab}$'s in favour
 of the ${\bf r}_a$'s, the quotient constriction eliminates the 
${ 1\over 2}m(m-1) $ phases of the $q_{ab}$'s so one use the $m$ quaternions
 $w_a$ as local coordinates on $X_\zeta$. Make the following
redefinitions:
\[ \mathbf{r}_a = \frac{1}{2}\mathbf{y}_a,\ \mathbf{x}_{ab} = 
\frac{1}{2}\boldsymbol{\zeta}_{ab}.\]
In these coordinates the metric is of the form (\ref{genmet}) with
potential functions given by:
\begin{eqnarray} \label{gibmanmet}
G_{aa} & = & 1+\sum_{b \ne a} \frac{1}{|\mathbf{r}_a - \mathbf{r}_b + 
             \mathbf{x}_{ab}|}\ (\mathrm{no\ sum\ over\ }a), \\
G_{ab} & = &-\frac{1}{|\mathbf{r}_a-\mathbf{r}_b+\mathbf{x}_{ab}|}. \nonumber
\end{eqnarray} 
The metric constructed by Gibbons and Manton in \cite{su2} is the
``negative mass'' version to obtain which one must take the flat
metric on $\mathcal{M}$ to be:
\begin{equation}
ds^2 = dq_{ab} d{\overline q} _{ab} - dw_a d {\overline w} _a
\end{equation}
and choose $\boldsymbol{\zeta} _{ab}=0$.
As pointed out in \cite{su2} the case of positive mass parameter appears to
be relevant to the motion of $a=1$ black holes. It also arises in
three-dimensional gauge theory. 
Physically the coordinates $w_a$  correspond to the positions and internal
 phases of the monopoles. Just as in the case of Lee-Weinberg-Yi more
complicated metrics may be constructed by introducing weights.

The construction is invariant under $m$ real translations of the $w_a$'s:
\begin{equation}
q_{ab} \rightarrow q_{ab}
\end {equation}
\begin{equation}
w_a \rightarrow w_a + t_a
\end{equation}
The global behaviour of these metrics is quite complicated, despite
the simplicity of the construction, and we hope to return to them in
a future publication. Note, as in the case of the Taubian-Calabi
metrics, the metric with $\boldsymbol{\zeta}_{ab}$ admit various
discrete symmetries, e.g. reflections and permutation groups, as
tri-holomorphic isometries. It follows that the $4m-$dimensional
metric contains totally geodesic copies of the first non-trivial case
$m=2$. This is presumably related to the observation of Bielawski that
the exact $SU(2)$ moduli space of $m$ monopoles always admits a
totally geodesic copy of the Atiyah-Hitchin manifold \cite{biel}.
\section{Zero- and Infinite-Mass Monopoles}  \label{massmon}
\subsection{Massless Monopoles}
In section~\ref{LWY} we showed how using \hk \ quotient construction one may
obtain the  exact metric on the relative moduli space of fundamental $SU(m+2)$ 
monopoles when the gauge group is maximally broken. It is interesting to ask
how the metric will change if some of the monopoles become massless, that is 
when the broken gauge group contains a non-abelian factor. This question was
addressed in a recent paper of Lee, Weinberg and Yi \cite{lwy2}. The argument 
used allowed the authors to obtain the metric for the case
\[ SU(m+2) \rightarrow SU(m)\times U(1)^2, \]
but did not yield an explicit answer for the more general case
\be
SU(m+2) \rightarrow SU(m+2-k) \times U(1)^k, 
\ee
$k = 2, \ldots,m+1$. Using the \hk \ quotient method, however, greatly
simplifies the task and we construct these metrics below. We will also
analyse  the metric on the moduli space of monopoles for the case $m=2, k=2$.
Let us first consider the case $m=1$.
\paragraph{Taub-NUT to flat metric:}
It is known \cite{conn,gaunt,lwy2} that the exact metric on the relative
moduli space of fundamental $SU(3)$ monopoles is the Taub-NUT metric with 
positive mass parameter. If one of the monopoles becomes massless 
the Taub-NUT metric degenerates to flat metric on $\re^{4}$.

In the notation of section~\ref{taubn} this is equivalent to
$\lambda \rightarrow \infty$. Define $\nu = \lambda^{-1}$, so $\nu 
\rightarrow 0$ when $\lambda \rightarrow \infty$. In order for the action 
(\ref{tnaction}) to be well defined we must introduce a new parameter 
$\ti{t}$ and a new quaternionic coordinate $\ti{w}$ such that:
\be  \label{newpar}
\nu \, \ti{t} = t, \ \ \ \ti{w} = \nu \, w.
\ee
Then the action (\ref{tnaction}) becomes:
\begin{eqnarray} \label{action}
q      & \rightarrow & q\, e^{i\nu \ti{t}}, \\
\ti{w} & \rightarrow & \ti{w} + \nu \ti{t}. \nonumber
\end{eqnarray}
In the limit $\nu \rightarrow 0$ the $U(1)$ action (\ref{action}) is
trivial and the metric on the quotient space $X_0$ is just the flat metric 
(\ref{flatmet}).
\paragraph{$\mathbf{SU(m+2) \rightarrow SU(m)\times U(1)^2}:$}
Physically this situation corresponds to having two massive monopoles and 
$m-1$ massless ones constituting a so-called massless cloud. We will
see that  in this
case the Lee-Weinberg-Yi metric degenerates to the Taubian-Calabi metric of 
section~\ref{taubcal}. 

When $m-1$ monopoles become massless, the reduced mass matrix $\mu_{ab}$ drops 
in rank to $\mathrm{rank} = 1$, which by (\ref{massmat}) is equivalent to 
$\nu_{ab}$ having rank one and the translation vectors $v^{(a)}$ not being 
linearly independent. By analogy with (\ref{newpar}) above we define new group 
parameters $\ti{t}_a$ and redefine quaternionic coordinates $w_a$ as:
\be
\nu_a^{\ b} \, \ti{t}_{b} = t_{a}, \  \ \ 
\ti{w}_a = \nu_a^{\ b} \, w_b.
\ee
Then  the action (\ref{lwyaction}) becomes:
\begin{eqnarray}  \label{newaction}
q_a      & \rightarrow & q_a\, e^{i\nu_a^{\ b} \ti{t}_b}, \\
\ti{w}_a & \rightarrow & \ti{w}_a + \nu_a^{\ b} \ti{t}_b. \nonumber
\end{eqnarray}
When the rank of $\nu_{ab}$ is one there is only one independent coordinate 
$\ti{w}_a$ and the $\re^m$ action (\ref{lwyaction}) reduces to the $\re$ 
action (\ref{newaction}). All elements of $\mu_{ab}$ are equal and the 
Lee-Weinberg-Yi metric (\ref{lwymet}) on $\re^{4m}$ degenerates to the 
Taubian-Calabi metric (\ref{tcal}) on $\re^{4m}$.
From section~\ref{taubcal} the tri-holomorphic part of the full isometry group
of the Taubian-Calabi metric that acts effectively is $U(m)/\mathbb{Z}_m$,
which agrees with the result of \cite{lwy3} up to a discrete factor.
In giving a physical interpretation to the degrees of freedom of the metric 
(\ref{tcal}) Lee, Weinberg and Yi pointed out that the massless monopoles 
cannot be regarded as individual particles. Instead they form a
so-called  massless 
cloud that carries no net non-abelian charge and is characterised by one 
``size'' parameter $R = \sum r_a$. This quantity is clearly invariant under 
the full isometry group $U(m)\times SU(2)$ since the metric on $\mathcal{M}$, 
and consequently $\sum q_a \bar{q}_a$, is preserved by both the $U(m)$ action
and the $SU(2)$ action, but $q_a \bar{q}_a = |q_a i \bar{q}_a| = r_a$. So $R$
is an invariant of the isometry group.

Let us focus on the simplest non-trivial case $m=2$. 
\paragraph{$\mathbf{SU(4)\rightarrow SU(2)\times U(1)^2}$:}
The Taubian-Calabi metric (\ref{tcal}) on $\re^8$ is:
\begin{eqnarray}  \label{su4}
4\,ds^2 & = &\Bigl(1+\frac{1}{r_1}\Bigr) d\mathbf{r}_1^2 + 2 d\mathbf{r}_1.
             d\mathbf{r}_2 + \Bigl(1+\frac{1}{r_2}\Bigr) d\mathbf{r}_2^2  \\
        &   & +\frac{1}{1+r_1+r_2} \Bigl[ r_1(1+r_2)(d\tau_1+\bs(\mathbf{r}_1)
               . d\mathbf{r}_1)^2 \nonumber \\
        &   & - 2 \,r_1 \,r_2 (d\tau_1+\bs
              (\mathbf{r}_1) . d\mathbf{r}_1)(d\tau_2+ \bs (\mathbf{r}_2)
              . d\mathbf{r}_2) +r_2(1+r_1)(d\tau_2+\bs(\mathbf{r}_2).
              d\mathbf{r}_2)^2 \Bigr],   \nonumber
\end{eqnarray}
where
\[\mathrm{curl} \bs_i = \mathrm{grad}\,\Bigl(\frac{1}{r_i}\Big),\ \ i = 1,2. \]
The action of $U(2)$ is tri-holomorphic and preserves $\mathbf{r}_1 + 
\mathbf{r}_2$, it has $4-$dimensional orbits. The left $SU(2)$ action preserves
the length of any vector and the inner products but rotates the
vectors around. 
Thus it rotates  $\mathbf{r}_1 + \mathbf{r}_2$ but keeps $|\mathbf{r}_1 + 
\mathbf{r}_2|$ invariant. This makes the principle orbits $6-$dimensional.
 
The metric (\ref{su4}) describes the moduli space of distinct centred 
fundamental $SU(4)$ monopoles in the limit when two of them become massless.
So we are left with two fundamental monopoles charged with respect to the two
$U(1)$'s and a ``massless'' cloud. Now there are $8$ parameters on the moduli
space: ${\mathbf{r}_1, \tau_1, \mathbf{r}_2, \tau_2}$. Four of them
correspond  to
position and phase of the two massive monopoles relative to centre of mass 
coordinates; so there are four parameters left to describe the cloud. 
Note that although we call the cloud ``massless'' it has non-zero
moment of inertia, in fact it has infinite moment of inertia. It has
zero  mass in the sense of a point 
particle, but cannot be regarded as such since it has a finite size. Infinite 
inertia means that it would take an infinite amount of energy to rotate this 
smeared out cloud. A good analogy to this situation is provided by 
$\mathbb{CP}^1$ lumps which are rational maps from $\mathbb{C}\cup
\{\infty\}$ onto $\mathbb{CP}^1$. The kinetic energy metric fails to
converge in some directions in moduli space because the lumps have
infinite moment of inertia \cite{lump}.

One may ask how the metric behaves if the cloud is large , i.e. if 
$|\mathbf{r}_1+\mathbf{r}_2|\gg |\mathbf{r}_1-\mathbf{r}_2|$, the
left-hand side is equivalent to $R$, the cloud size parameter. Since
swapping $q_1$ and $q_2$ is an isometry that preserves the \hk\
structure, the submanifold $q_1 = q_2$ of the fixed points of this
isometry, which implies $\mathbf{r}_1 = \mathbf{r}_2$ and $\tau_1 =
\tau_2$, is totally geodesic. In fact, it is isomorphic to the
Taub-NUT space. Therefore for large separations the cloud part of the
metric, whose coordinates are $(\mathbf{r}_1 + \mathbf{r}_2)$ and
$\frac{1}{2}(\tau_1+\tau_2)$, behaves like the Taub-NUT metric. This
is consistent with the fact that the cloud has four degrees of freedom
-- the position of its centre of mass and a phase. It has no
rotational degrees of freedom, which is consistent with the claim that
its moment of inertia is infinite. 
\paragraph{$\mathbf{SU(m+2)\rightarrow SU(m+2-k)\times U(1)^k}:$}
We can now construct all the intermediate cases when the broken gauge
group $SU(m+2)$ contains one non-abelian factor. These can be viewed
as generalisations
of the Taubian-Calabi metrics (\ref{tcal}) with the initial \hk \ manifold now 
being
\[ \mathcal{M} = \q^m \times \q^{k-1}. \]
Now $(m+1-k)$ monopoles become massless, $k = 2,\ldots,m$, so rank of 
$\mu_{ab}$ and rank of $\nu_{ab}$ is equal to $k-1$. Looking at the action 
(\ref{newaction}) we can see that there are only $k-1$ independent coordinates 
$\ti{w}_i$, $i = 1,\ldots,k-1$. The potential function $G_{ab}$ is the same as 
in (\ref{lwymet}) with $\mu_{ab}$ of the form:
\[ \mu_{ab} = \Biggl( \begin{array}{c|c}
                        \mu_{ab}' & \ldots \\
                        \noalign{\hrule} 
                        \vdots & \ddots
                       \end{array} \Biggr) \]
where $\mu_{ab}'$ is the $(k-1)\times (k-1)$ reduced mass matrix for $k$ 
fundamental monopoles with the rest of the entries all equal. It is not 
difficult to see that the isometry group for this manifold up to a 
discrete factor will be:
\[ SU(2)\times U(m-(k-2))\times U(1)^{k-2}.\]
\subsection{Infinitely massive monopoles}
Here we will consider another interesting degeneration of the Lee-Weinberg-Yi
metric which occurs when the translation matrix $\lambda_{ab}$ drops in rank.
Let $\lambda_{ab}$ have rank one. Then the $\re^m$ action (\ref{lwyaction})
\begin{eqnarray} 
q_a & \rightarrow & q_a \, e^{it_a} \ \ \   (\mathrm{no\  sum\  over}\ a) \\
w_a & \rightarrow & w_a + \lambda_a^{\ b} t_b   \nonumber
\end{eqnarray}
reduces to the $\re^m$ action (\ref{alfmm}), and there is effectively one $w_a$ 
coordinate. The initial manifold $\mathcal{M}$ is now $\q^m \times \q$ and the
setup is equivalent to the setup for cyclic ALF spaces in section~\ref{alf}. 
Physically this represents situation where moduli space coordinates of all 
but one monopoles are fixed (all except one monopoles are infinitely heavy) 
and one monopole (described by the one $w_a$ coordinate) moves in their 
background. We get the multi-centre metric with ALF boundary 
conditions (\ref{alfmet}).

If the rank of $\lambda_{ab}$ is $1< k < m$, we get a $4k-$dimensional
generalisation of cyclic ALF spaces.
\section{Geodesics and Integrability}  \label{geodesic}
The slow {\it classical} motion of monopoles corresponds to geodesics on the
moduli space \cite{manton}. It is of interest to know whether there exist any
closed or bound\footnote{By bound we mean confined to a compact set for 
all times.} geodesics which would describe (classical) bound states of 
monopoles. Note that here we are not talking about the zero-energy
threshold bound states predicted by Sen's conjecture, but about true bound
states with positive energy.

If the topology of the moduli space is complicated one may invoke a general
result of Benci and Giannoni \cite{benci} for open manifolds to establish 
existence of closed geodesics. However if the manifold is
topologically  trivial, such arguments give no information.

For topologically trivial manifolds such as Lee-Weinberg-Yi and Taubian-Calabi
spaces one may use the following criteria. If there exists an everywhere
distance increasing vector field $\mathbf{V}$ then there are no closed or 
bounded geodesics on this manifold. Distance increasing condition means that
the Lie derivative of the  metric along $\mathbf{V}$ satisfies:
\be
\mathcal{L}_{\scriptscriptstyle \mathbf{V}} g(\mathbf{X},\mathbf{Y}) > 0,
\ee
for all $4m-$vectors $\mathbf{X},\mathbf{Y}$, or equivalently 
\[ \mathbf{V}_{(a;b)}\mathbf{X}^a \mathbf{Y}^b >0. \]   
Along a geodesic with a tangent vector $\mathbf{L}$ one therefore has:
\be  \label{geod}
{d\over dt}  g(\mathbf{V},\mathbf{L}) =\mathcal{L}_{\scriptscriptstyle 
\mathbf{V}} g(\mathbf{L},\mathbf{L}) > 0.
\ee
Now if this is a bound or closed  geodesic one may average over a time period
$T$. The left-hand side of (\ref{geod}) tends to zero as $T \rightarrow\infty$
while the right-hand side tends to some positive constant, which is a 
contradiction.

The existence of a distance increasing vector field can be easily demonstrated
on spaces obtained by \hk \ quotient restricting to zero-set of the moment map
\footnote{It will be clear shortly why this argument does not apply to spaces
where one cannot without loss of generality consider $\mm{0}$.}. From our 
examples in section~\ref{example} these are Lee-Weinberg-Yi and Taubian-Calabi
manifolds. The vector field $\mathbf{V}$ is induced on $X_0$ from the 
following $\re^+$ action on $\mathcal{M} = \q^m \times \q^p$:
\begin{eqnarray} \label{homothety}
q_a & \rightarrow & \alpha^{1/2} q_a \\
w_i & \rightarrow & \alpha \, w_i , \ \ \  \alpha > 0 \nonumber
\end{eqnarray}
$a = 1, \ldots, m$ and $i = 1,\ldots,p$. This $\re^+$ action leaves invariant 
the level sets $\mm{0}$ and commutes with the $\re^m$, $T^m$ and $SU(2)$ 
actions. It therefore descends to give a well-defined $\re^+$ action on 
$\mm{0}/\re^m$ which stabilises the point $q_a = 0$ corresponding for 
Lee-Weinberg-Yi metric to the spherically symmetric monopole \cite{lwy2}. The 
action (\ref{homothety}) is clearly distance increasing on $\mathcal{M}$, 
so its restriction to $\mm{0}/\re^m$ is also distance increasing. Note
that the argument just given is a more geometric version of the
generalised Virial Theorem given earlier in \cite{gary}.

We conclude this section by making a remark about the integrability of the 
geodesic flow. Consider the Lagrangian for the configuration described
by the moduli space metric (\ref{genmet}) (see \cite{su2}). If one
eliminates the conserved charges
\[ Q^a = G^{ab} \Bigl( \frac{d\tau_a}{dt} + \bs_{ac} .\frac{d\mathbf{r}_c}
{dt}\Bigr), \]
one obtains an effective Lagrangian on $\re^{3m} = X_\zeta /T^m$ :
\be  \label{lagran}
G_{ab} \frac{d\mathbf{r}_a}{dt} . \frac{d\mathbf{r}_b}{dt} - G_{ab}\,Q^a 
\,Q^b + Q^a \bs_{ac} . \frac{d\mathbf{r}_c}{dt}.
\ee
This many-body Lagrangian (\ref{lagran}) may not look very tractable when 
considered on $\re^{3m}$ but in some cases it admits ``hidden'' symmetries 
which, although not apparent in $3m$ dimensions, are clearly present on the 
$4m-$dimensional manifold. A simple example of the phenomenon occurs in the 
Eguchi-Hanson manifold. On $\re^3$ there is only one manifest symmetry 
corresponding to rotation about the axis joining the two centres. However, the
geodesic motion is completely integrable \cite{mignemi}. This happens because 
of the large
isometry group, $U(2)$, which acts on the three-dimensional orbits. In
fact the motion
on electrically neutral geodesics, i.e. those with $Q=0$, in the  ALE and ALF 
spaces associated to the cyclic group of order $k$ is the same as that of a 
light planet moving in the Newtonian gravitational field of $k$ fixed 
gravitating centres. The case $k=1$ corresponds to the Kepler problem and is 
clearly integrable, The case $k=2$ corresponds to the Euler problem and is also
integrable. According to \cite{fomenko} the case when there is a plane 
containing the centres and the forces are attractive and the motion of a 
planet with positive energy is confined to that plane, then there are no  
analytic constants of the motion other than the energy if $k>2$. Since this
case is a special case of the general motion, it strongly indicates that for 
$k>2$ the geodesic flow on the cyclic ALE and ALF spaces is not integrable. 
One might be able to use a similar argument in some other cases.
   
However  one might expect to encounter hidden symmetries in the case of the 
Calabi and the Taubian-Calabi  metrics. As we have seen above neither the 
Calabi nor the Taubian-Calabi metrics  look very symmetric when written out 
in terms of the cartesian coordinates but in fact they admit a large group of 
isometries whose principal orbits are of co-dimension one or two respectively. 
Another interesting question along the same lines is whether there are cases 
in which the geodesic flow admits Lagrange-Laplace-Runge-Lenz vectors as it
does in the Taub-NUT case. We defer further considerations of these questions
for a future publication.
\section{Concluding Comments}
In this paper we have, using the \hk\ quotient, presented a rather simple
and elegant way to construct and analyse some known and some new \hk\
metrics. All our examples turned out to possess a tri-holomorphic
torus action which considerably simplified the algebra. From the few
applications that we have discussed, it is clear that this approach
gives explicit answers to many interesting questions about the global
properties of these manifolds. In many cases such properties are not
immediately apparent from the local form of the metric. 

There are a number of open problems that can be explored using the \hk\
quotient.   
In the future we hope to extend the method introduced in this paper further to
discuss the differential forms and the spectrum of the Hodge-de Rahm Laplacian,
as well as the physics of so-called massless monopoles. It would also
be interesting to look at the singular metrics and understand the type
of singularities that arise and how they may be resolved.
\subsection*{Acknowledgements}
We would like to thank R.Goto and N.Hitchin for illuminating
conversations. The research of P.R. is supported by Trinity College,
Cambridge University. 

\begin{thebibliography}{99}
\bibitem{hkquo}N.J.Hitchin, A.Karlhede, U.Lindstr\"{o}m and M.Ro\v{c}ek, 
{\em \hk \ metrics and Supersymmetry}, Comm.Math.Phys. {\bf 108} (1987) 535.
\bibitem{lwy1}K.Lee, E.J.Weinberg and P.Yi, {\em The Moduli Space of Many BPS
Monopoles}, hep-th/9602167.
\bibitem{gaunt}J.P.Gauntlett and D.A.Lowe, {\em Dyons and S-duality in $N=4$ 
Supersymmetric Gauge Theory}, hep-th/9601085.
\bibitem{lwy2} K.Lee, E.J.Weinberg and P.Yi, {\em Electromagnetic Duality and 
$SU(3)$ Monopoles}, hep-th/9601097.
\bibitem{gary}G.W.Gibbons, Phys.Lett. {\bf 382B} (1996) 93.
\bibitem{intril}K.Intriligator and N.Seiberg, {\em Mirror Symmetry in
Three Dimensional Gauge Theories}, hep-th/9607207.
\bibitem{su2}G.W.Gibbons and N.S.Manton, {\em The moduli space metric for 
well-separated BPS monopoles}, Phys.Lett.{\bf B356} (1995) 32.
\bibitem{lwy3}K.Lee, E.J.Weinberg and P.Yi, {\em Massive and Massless 
Monopoles with Nonabelian Magnetic Charge}, hep-th/9605229.
\bibitem{linroc}U.Lindstr\"{o}m and M.Ro\v{c}ek, Nucl.Phys. {\bf B222}
(1983) 285.
\bibitem{poon}H.Pedersen and Y.S.Poon, Comm.Math.Phys. {\bf 117} (1988) 569.
\bibitem{goto}R.Goto, {\em On Toric \hk\ Manifolds}, Kyoto University
preprint RIMS$-818$ (1991); {\em On \hk\ Manifolds of Type
$A_{\infty}$}, Geom.Funct.Anal. {\bf 4(4)} (1994) 424 
\bibitem{mur}M.K.Murray, {\em A Note on the $(1,1,\ldots,1)$ Monopole 
Metric}, hep-th/9605054.
\bibitem{conn}S.A.Connell, {\em The Dynamics of the $SU(3)$ charge $(1,1)$
Magnetic Monopoles}, University of South Australia preprint.
\bibitem{calabi}T.L.Curtright and D.Z.Freedman, Phys.Lett. {\bf 90B} (1980) 71.
\bibitem{swan}A.Dancer and A.Swann, {\em \hk \ Metrics of Cohomogeneity One}, 
J.Geom.Phys. to appear.
\bibitem{tcalabi}M.Ro\v{c}ek, Physica {\bf 15D} (1985) 75 .
\bibitem{minst}G.W.Gibbons and S.W.Hawking, Phys.Lett.{\bf B78(4)} (1978) 430.
\bibitem{mtn}S.W.Hawking, Phys.Lett.{\bf 60A} (1977) 81.
\bibitem{biel}R.Bielawski, {\em Monopoles, Particles and Rational
Functions}, McMaster University preprint (1994).
\bibitem{lump}R.S.Ward, Phys.Lett. {\bf 158B} (1985) 424.
\bibitem{manton}N.S.Manton, Phys.Lett.{\bf B110} (1982) 54.
\bibitem{benci}V.Benci and F.Giannoni, Duke Math.J. {\bf 68(2)} (1992) 195.
\bibitem{mignemi}S.Mignemi,{\em Classical and Quantum Motion on an 
Eguchi-Hanson Space}, J.Math.Phys.{\bf 32(11)} (1991) 3047 .
\bibitem{fomenko}A.T.Fomenko, {\em Symplectic Geometry}, Advanced Studies in
Contemporary Mathematics {\bf 5}, Gordon and Breach (1988).
%
%
%
\end{thebibliography}
\end{document}